\documentclass[a4paper,12pt]{article}

\usepackage[a4paper,text={16.8cm,22.8cm}]{geometry}
\usepackage{amsmath,amssymb,bm,graphicx,color}
\usepackage{extarrows}
\usepackage[utf8]{inputenc}
\usepackage{amsthm}
\usepackage{slashed}
\usepackage{tikz}
\usepackage{rotating}
\usepackage{hyperref}
\usepackage{array}
\usepackage{color}
\usepackage{multirow}
\usepackage{bbold}
\usepackage{cite}
\usepackage{pifont}
\usepackage{tabularx}

\allowdisplaybreaks 
\renewcommand{\arraystretch}{1.2}

\begin{document}

\begin{titlepage}

\vspace{1.2cm}
\begin{center}
\Large\bf
\boldmath
The Geometric $\nu$SMEFT:  Operators and Connections
\unboldmath
\end{center}
\vspace{0.2cm}
\begin{center}
{\large{Jim Talbert$^a$}}\\
\vspace{1.0cm}
{\sl 
${}^a$\,\small DAMTP, University of Cambridge, Wilberforce Rd., Cambridge, CB3 0WA, United Kingdom}\\[0.5cm]
{\bf{E-mail}}:  rjt89@cam.ac.uk
\end{center}

\vspace{0.5cm}
\begin{abstract}
\vspace{0.2cm}
\noindent 
We write down a geometric realization of the Standard Model Effective Field Theory (SMEFT) extended by $n_f$ flavours of light sterile neutrinos, a so-called geo$\nu$SMEFT.  As with the geoSMEFT introduced by Helset, Martin and Trott, we show that a refactorization of the $\nu$SMEFT's operator product expansion is possible, such that two- and three-point composite operator forms are dressed with field-space connections composed of towers of Higgs dressings and symmetry generators, valid at \emph{all-orders} in the $\overline{v}_T/\Lambda$ expansion parameter of the EFT ($\overline{v}_T \equiv \sqrt{2\langle H^\dagger H\rangle}$) .  These connections are parameterized by real Higgs coordinates and contribute to the field-space geometry of the ($\nu$)SM, with structure  linked to the strength of Beyond-the-($\nu$)Standard Model physics encoded in $\overline{v}_T/\Lambda$. In addition to enumerating the relevant composite operators and associated connections, we briefly outline the route to calculating all-$\overline{v}_T/\Lambda$-orders amplitudes, including the flavor-invariant theory required to understand the neutrino mass-eigenstate basis geometrically.
\end{abstract}
\vfil

\end{titlepage}


\tableofcontents
\noindent \makebox[\linewidth]{\rule{16.8cm}{.4pt}}


\section{Introduction and Motivation}
\label{sec:INTRO}
Neutrino physics represents an ideal sector for probing novel interactions Beyond-the-Standard Model (BSM).  After all, the very presence of non-zero neutrino masses and Pontecorvo-Maki-Nakagawa-Sakata (PMNS) mixings, as unambiguously inferred from global oscillation experiments (cf. \cite{Esteban:2020cvm}), requires the introduction of operators beyond those furnished by the (renormalizable) SM Lagrangian.  For example, upon allowing for an operator-product-expansion (OPE) in \emph{non}-renormalizable interactions, the SM Effective Field Theory (SMEFT) can easily generate a Majorana neutrino mass term via the dimension-five Weinberg Operator \cite{Weinberg:1979sa} $\mathcal{Q}_5$,
\begin{equation}
\label{eq:SMEFTwithWeinberg}
\mathcal{L}_{\text{SMEFT}} \equiv \mathcal{L}_{\text{SM}} + \sum_{i}\tilde{C}_i \mathcal{Q}_i = \mathcal{L}_{\text{SM}} + \frac{1}{2\Lambda}\left[C_{\underset{pr}{5}}\left(\tilde{H}^\dagger \ell_p \right)^T \mathbb{C}\left( \tilde{H}^\dagger \ell_r\right) + \text{h.c.}\right] + \mathcal{O}(1/\Lambda^2) + ...\,,
\end{equation}
upon electroweak symmetry breaking, when the scalar Higgs field $H$ acquires a vacuum expectation value (VEV) $\overline{v}_T \equiv \sqrt{2\langle H^\dagger H \rangle}$.  Here $C_{i}$ represent unknown Wilson Coefficients (with $p,r$ flavor labels) that parameterize the infrared (IR) effects on local contact interactions $\mathcal{Q}_i$ coming from unspecified, decoupled  ultraviolet (UV) dynamics propagating at an arbitrary new physics scale $\Lambda$, with $\overline{v}_T/\Lambda << 1$. $\mathbb{C}$ is the charge conjugation matrix and $\tilde{H}$ is given in terms of the Levi-Civita  tensor $\epsilon$, $\tilde{H}_j \equiv \epsilon_{jk} H^{\dagger k}$.\footnote{Note that our convention for $\epsilon$ is  $\epsilon^{12}=-\epsilon^{21}=\epsilon_{21}=-\epsilon_{12}=-1$.}  The SMEFT is therefore composed of all (non-)renormalizable operators $\mathcal{Q}^{(d)}$ of mass dimension $d$, invariant under spacetime and  SU(3)$_c \times$ SU(2)$_L \times$ U(1)$_Y$ gauge symmetries of the SM, composed of SM fields only (cf. the first seven columns of Table \ref{tab:nuSMEFTfieldandsymm}), with unspecified couplings/coefficients $C_i^{(d)}$.

Similar in spirit to the SMEFT, the $\nu$SMEFT allows for all (non-)renormalizable gauge- and spacetime-invariant operators composed of SM fields, but also introduces $n_f$ flavors of a light gauge singlet neutrino $N$ into the IR spectrum.\footnote{Note that $N$ is the standard right-handed (RH), four-component neutrino spinor sometimes denoted by $N_R$ in the literature (see e.g. \cite{Broncano:2002rw}).  We have removed the $R$ subscript for clarity of notation, and for consistency with the final results of \cite{Li:2021tsq}.  Also, $N^c \equiv \mathbb{C} \overline{N}^T$.}  The field and gauge symmetry content of the $\nu$SM(EFT) is also presented in Table \ref{tab:nuSMEFTfieldandsymm}, while its Lagrangian is given by
\begin{equation}
\label{eq:nuSMEFT}
    \mathcal{L}_{\nu\text{SMEFT}} \equiv \mathcal{L}_{\text{SM}} + \mathcal{L}_N + \sum_i \frac{C_i^{(d)}}{\Lambda^{d-4}} \mathcal{Q}_i^{(d)} \,,
\end{equation}
where the renormalizable interactions of $N$ are simply 
\begin{equation}
\label{eq:LagSS}
\mathcal{L}_N = \overline{N}\,i\slashed{\partial}\,N - \frac{1}{2}\left[ \overline{N}\,M\,N^c + \overline{N^c}\,M^\star\,N\right] -  \overline{\ell_L}\, \tilde{H} \, Y_N \, N - \overline{N}\,Y_N^{\dagger}\,\tilde{H}^\dagger\,\ell_L \,,
\end{equation}
which respectively constitute a kinetic term, lepton-number-violating (LNV) Majorana mass terms, and novel Yukawa interactions that serve as a portal to the SM.  Note also that in \eqref{eq:nuSMEFT} we have included the relevant scale suppression factor $\Lambda^{d-4}$ associated to a given mass-dimension-$d$ operator.  Hence the $\nu$SMEFT is the relevant EFT describing nature at energy scales $M_{\nu\text{SM}} << \Lambda$, such that \eqref{eq:nuSMEFT} allows one to calculate neutrino amplitudes in a model-independent way, encoding the effects of heavier particles associated to explicit new physics scenarios into (potentially) non-zero Wilson Coefficients.  Indeed, accounting for the possibility of $N$-dependent IR vertices is well-motivated not only by the observation of non-zero light neutrino masses, but also by deeper theoretical concerns about the ability to embed the SM enhanced by three Majorana neutrinos into a quantum theory of gravity \cite{Arkani-Hamed:2007ryu,Ooguri:2016pdq,Ibanez:2017kvh,Gonzalo:2021zsp}, as well as a host of experimental anomalies in short-baseline neutrino oscillation experiments dating back to results from the Los Alamos Liquid Scintillator Neutrino Detector (LSND) --- see e.g. \cite{Dasgupta:2021ies} for a recent review on sterile neutrino physics.  As a result, significant theoretical interest has developed around \eqref{eq:nuSMEFT}: building upon earlier results in \cite{delAguila:2008ir,Aparici:2009fh,Bhattacharya:2015vja,Liao:2016qyd}, a complete operator basis up to $d = 9$ is now available from \cite{Li:2021tsq}, as are matching and renormalization group (RGE) analyses at tree and one-loop accuracy \cite{Chala:2020vqp,Chala:2020pbn}, along with numerous phenomenological studies at low and high energies (see e.g. \cite{Duarte:2015iba,Duarte:2016caz,Ballett:2016opr,Cai:2017mow,Alcaide:2019pnf,Butterworth:2019iff,Bischer:2019ttk,Dekens:2020ttz,Chala:2020pbn,Bolton:2020xsm,Duarte:2020vgj,Biekotter:2020tbd}).

\begin{table}
\centering
{\renewcommand{\arraystretch}{1.4}
\begin{tabular}{|c||c|c|c|c|c|c|c|}
\hline
& $H$ & $q_L$ & $\ell_L$ & $u_R$ & $d_R$ & $e_R$ & $N$ \\
\hline
\hline
SU(3)$_c$ & {\bf{1}} & {\bf{3}} & {\bf{1}} & {\bf{3}} & {\bf{3}} & {\bf{1}} & {\bf{1}} \\
\hline
SU(2)$_L$ & {\bf{2}} & {\bf{2}} & {\bf{2}} & {\bf{1}} & {\bf{1}} & {\bf{1}} & {\bf{1}} \\
\hline
U(1)$_Y$ & $\frac{1}{2}$ & $\frac{1}{6}$ & $-\frac{1}{2}$ & $\frac{2}{3}$ & $-\frac{1}{3}$ & $-1$ & 0 \\
\hline
\end{tabular}}
\caption{
Field and (gauge) symmetry content of the $\nu$SM(EFT).}
\label{tab:nuSMEFTfieldandsymm}
\end{table}
Concurrent to growing interest and developments with the $\nu$SMEFT, recent progress has also been made in understanding the mathematical behavior associated to arbitrarily high numbers of (non-derivative) Higgs interactions in generic EFTs, parameterized by (e.g.) successive $H^\dagger H$ insertions into the EFT Lagrangian.  At arbitrary mass-dimensions these scalar dressings generalize into well-defined \emph{field-space connections}, compact objects that define geometries (e.g. metrics) on the field spaces defined by associated classes of EFT operator forms.  Critically, they can be defined at \emph{all-orders} in the $\overline{v}_T/\Lambda$ expansion of the (e.g.) SMEFT, are in many instances field-redefinition invariant and, at least at tree level, can be used to reabsorb an arbitrary tower of Wilson coefficients into a single mathematical object, thereby reducing the free parameters of the EFT.  Exploiting this fact, a so-called geometric SMEFT (geoSMEFT) \cite{Helset:2020yio} has been developed recently, drawing on prior geometric insights into the scalar sector of generalized Higgs Effective Field Theory \cite{Alonso:2015fsp,Alonso:2016oah}, as well as developments in gauge-fixing \cite{Helset:2018fgq} and demonstrating Ward identities \cite{Corbett:2019cwl} in the standard formulation of the SMEFT.  The geoSMEFT imposes a refactorization on the SMEFT's OPE, such that
\begin{equation}
\label{eq:geofactorize}
    \mathcal{L}_{\text{SMEFT}} \overset{!}{\equiv} \sum_{i} G_i \left(I, A, \phi, ...\right)\, f_i \,,
\end{equation}
where $G_i$ are the aforementioned field-space connections which depend on the real coordinates $\phi$ of the SM Higgs doublet parameterized as
\begin{equation}
    \label{eq:RealHiggs}
    H(\phi_I) = \frac{1}{\sqrt{2}} \left[
    \begin{array}{c}
    \phi_2 + i \phi_1 \\
    \phi_4 - i \phi_3
    \end{array}
    \right],\,\,\,\,\,\,\,\,\,\,
    \tilde{H}(\phi_I) = \frac{1}{\sqrt{2}} \left[
    \begin{array}{c}
    \phi_4 + i \phi_3 \\
    -\phi_2 + i \phi_1
    \end{array}
    \right],
\end{equation}
as well as (potentially) group indices $I$, $A$ associated to symmetry generators of the SM gauge group, while the \emph{composite operator forms} $f_i$ carry all of the non-trivial Lorentz indices of the SMEFT, including gauge-covariant derivatives of the Higgs field, $D^\mu H$.  For example, in the gauge sector of the geo($\nu$)SMEFT one can readily identify the composite operator $f_{\mathcal{W}\mathcal{W}} = \mathcal{W}_{\mu\nu}^A \mathcal{W}^{B,\mu\nu}$, where the indices $A,B$ run over the four electroweak gauge bosons (see Section \ref{sec:OPERATORS} below), and its associated connection $G_{\mathcal{W}\mathcal{W}} = g_{AB}(\phi)$ \cite{Helset:2020yio},\footnote{Here $\Gamma$ are combinations of electroweak symmetry generators written in a four-dimensional real representation (cf. Appendix \ref{sec:CONVENTIONS}) and $\tilde{C}_i$ represent the standard notation for Wilson coefficients (normalized to the new physics scale $\Lambda$) on effective operators in the gauge-Higgs sector of the SMEFT, e.g. $\mathcal{L}_{SMEFT} \supset \tilde{C}_{HB}^{(6+2n)} (H^\dagger H)^{(n+1)} B^{\mu\nu}B_{\mu\nu}$ --- see \cite{Helset:2020yio} for more details.}
\begin{align}
\nonumber
    g_{AB}(\phi) &= \left[1- 4 \sum_{n=0}^{\infty} \left(\tilde{C}_{HW}^{(6+2n)} \left(1-\delta_{A4}\right) + \tilde{C}_{HB}^{(6+2n)} \delta_{A4}
    \right) \left(\frac{\phi^2}{2}\right)^{n+1}\right] \delta_{AB} \\
    \nonumber
    &-\sum_{n=0}^{\infty}\left[ \tilde{C}_{HW,2}^{(8+2n)} \left(\phi_I \Gamma_{A,J}^I \phi^J \right) \left(\phi_L \Gamma_{B,K}^L \phi^K \right) \left(1-\delta_{A4} \right) \left(1-\delta_{B4}\right)\right]\left(\frac{\phi^2}{2}\right)^n  \\
    \label{eq:gaugeconnection}
    &+\sum_{n=0}^{\infty}\tilde{C}_{HWB}^{(6+2n)}\left[ \left(\phi_I \Gamma_{A,J}^I \phi^J \right) \left(1-\delta_{A4} \right) \delta_{B4} + \left(A \leftrightarrow B \right)\right]\left(\frac{\phi^2}{2}\right)^n\,,
\end{align}
which is clearly defined at all $\overline{v}_T/\Lambda$ orders, and which also amounts to a metric on the field space defined by $f_{\mathcal{W}\mathcal{W}}$;  sending the Wilson coefficients $\tilde{C}$ to zero recovers the (diagonal) SM limit, and hence the size of new physics embedded in $\tilde{C}$ determines the curvature of this field space. One can also obtain analytic forms for the all-orders generalizations of important theory parameters from the definitions of $G_i$ like $g_{AB}$, e.g. gauge and Higgs boson masses, gauge couplings, weak mixing angles \cite{Helset:2020yio}, and even the fermionic mass and mixing angles of the Dirac Yukawa (flavor) sector of the (geo)SM(EFT) \cite{Talbert:2021iqn}. The geoSMEFT also gives rapid insight into the applicability/validity of schemes for estimating uncertainties coming from higher-order effects in fixed-order phenomenological SMEFT studies \cite{Hays:2020scx,Corbett:2021cil}, has facilitated a global fit to electroweak precision data \cite{Corbett:2021eux} and dedicated  mono-lepton production study in the $d=8$ SMEFT \cite{Kim:2022amu}, and is also applicable to loop calculations as demonstrated in \cite{Corbett:2021jox} for the SMEFT's Higgs tadpole. 

Hence both the $\nu$SMEFT and geoSMEFT amount to promising effective theories with significant phenomenological applicability to current and future experiment.  In what follows we aim to merge these two technologies by presenting the first `geometric' realization of the $\nu$SMEFT --- a so-called \emph{geo$\nu$SMEFT}.  We do so by enumerating the composite operator forms $f_i$ associated to two- and three-point vertices in the geo$\nu$SMEFT in Section \ref{sec:OPERATORS}, while in Section \ref{sec:CONNECTIONS} we define the all-orders field space connections $G_i$ associated to the saturation of $f_i$ in mass dimension $d$, which we demonstrate using (automated) Hilbert Series techniques.  In Section \ref{sec:PHENO} we then briefly discuss calculating all-orders amplitudes in the geo$\nu$SMEFT, including the flavor-invariant theory required to describe the transformations to its neutrino mass-eigenstate basis in a geometric way. We then conclude in Section \ref{sec:CONCLUDE}, and collect some conventions and definitions required to reproduce our results and better understand the geometric formalism in Appendix \ref{sec:CONVENTIONS}.

\section{geo$\nu$SMEFT Composite Operator Forms}
\label{sec:OPERATORS}

\begin{table}
\centering
{\renewcommand{\arraystretch}{1.4}
\begin{tabular}{|c||c|c|c|c|c|c|c|}
\hline
\multicolumn{8}{|c|}{$\nu$SMEFT Operator Counting} \\
\hline
Mass Dimension & 5 & 6 & 7 & 8 & 9 & 10 & 11 \\
\hline
\hline
\multirow{2}{*}{$n_f = 1$} & 4 & 113 & 110 & 1316 & 1918 & 21540 & 37354 \\
& 2 & 29 & 80 & 323 & 1358 & 6084 & 25392\\
\hline
\multirow{2}{*}{$n_f = 2$} & 14 & 1037 & 1226 & 14008 & 41720 & 435452 & 1191386 \\
 & 8 & 343 & 894 & 4205 & 30102 & 160805 & 820964\\
\hline
\multirow{2}{*}{$n_f = 3$}  & 30 & 4659 & 5748 & 65207 & 334400 & 3513704 & 11347838 \\
 & 18 & 1614 & 4206 & 20400 & 243944 & 1421263 & 7875572 \\
\hline
\end{tabular}}
\caption{
Operator counting up to mass dimension 11 in the $\nu$SMEFT with one, two, and three flavors $n_f$, assuming that the number of sterile neutrino flavors is equal to the number of SM fermion flavors.  This table is generated with {\tt{ECO}} \cite{Marinissen:2020jmb} and can be trivially extended to higher mass dimensions.  Note that our counting on the top row, for a given number of flavors, includes both the operators of the traditional SMEFT as well as those with $N$ dependence, while the counting on the second row only includes the number of novel $N$-dependent operators at said mass dimension. From this one sees that at $n_f=1\, (3)$ we obtain 84 (3045) independent SMEFT operators at mass dimension six, in line with standard counting schemes.
\label{tab:nuSMEFTcounting}
}
\end{table}

Constructing the geo$\nu$SMEFT amounts to imposing the geometric factorization of \eqref{eq:geofactorize} on the traditional formulation of  the $\nu$SMEFT Lagrangian,
\begin{equation}
\label{eq:nugeofactorize}
    \mathcal{L}_{\nu\text{SMEFT}} \overset{!}{\equiv} \sum_{i} G_i \left(I, A, \phi, ...\right)\, f_i \,,
\end{equation}
and so our principal task is to identify the $G_i$ and $f_i$ in this theory.  In this Section we approach the latter composite operator forms $f_i$, recalling that \cite{Helset:2020yio}
\begin{itemize}
    \item whilst a geometric formulation of an EFT is possible regardless of operator basis, the exact analytic forms of the $f_i$ and corresponding $G_i$ can take on a basis dependence.
    \item a \emph{finite} list of $f_i$ can only be found for two- and three-point $f_i$, where a `point' can constitute a field-strength tensor $X_{\mu\nu}$, a fermion $\psi$, a Higgs-derivative term $D^\mu \phi$, or a fermion-derivative term $D^\mu \psi$.  This is due to the failure of integration-by-parts identities to reduce out higher-derivative operators acting on ($n>3$)-point functions.
\end{itemize}
With this in mind we will construct a geometric formulation of the two- and three-point $\nu$SMEFT that is consistent with the operator basis presented in \cite{Li:2021tsq} up to $d=9$, which is simultaneously consistent with the output of a Hilbert Series calculation.\footnote{We have explicitly checked that the number of operators presented in \cite{Li:2021tsq} is consistent with Table \ref{tab:nuSMEFTcounting}.  For example, \cite{Li:2021tsq} finds that there are $\mathcal{N}(n_f)=n_f/72 \, (16651 \,n_f^5 + 327\, n_f^4 + 64519 \,n_f^3 - 1335 \,n_f^2 + 17182\, n_f + 432)$ operators in the $\nu$SMEFT at $d=9$, which yields $\lbrace \mathcal{N}(1), \mathcal{N}(2), \mathcal{N}(3) \rbrace = \lbrace 1358, 30102, 243944 \rbrace$.}  Our notation for $f_i$ will follow the four-component spinor notation presented there, and will otherwise also respect the conventions presented in \cite{Helset:2020yio}.

To the latter end we follow \cite{Helset:2020yio} and, in addition to the real scalar field coordinates $\phi_I = \lbrace \phi_1, \phi_2, \phi_3, \phi_4 \rbrace$ parameterizing \eqref{eq:RealHiggs}, we also combine the SM electroweak gauge bosons and corresponding couplings into four-vectors $\mathcal{W}_A = \lbrace W_1, W_2, W_3, B \rbrace$ and $\alpha_A = \lbrace g_2, g_2, g_2, g_1 \rbrace$.  Noting that the bosonic field content of the $\nu$SMEFT is identical to that of the SMEFT, we can then define the transformations to the bosonic mass-eigenstate basis via
\begin{equation}
\label{eq:bosonbasischange}
    \mathcal{U}_C^A = \sqrt{g}^{AB} U_{BC}\,\,\,\,\,\,\,\,\,\text{and}\,\,\,\,\,\,\,\,\,\, \mathcal{V}^I_K = \sqrt{h}^{IJ} V_{JK}\,,
\end{equation}
where $\sqrt{g}^{AB}$ ($\sqrt{h}^{IJ}$) is the matrix square-root of the expectation value of \eqref{eq:gaugeconnection}, the field-space connection associated to the $\mathcal{W}^A\mathcal{W}^B$ ($(D_\mu \phi)^I(D_\mu \phi)^J$) bilinear composite operator in the standard geoSMEFT, $\mathcal{L} \supset g_{AB}(\phi) \mathcal{W}^A \mathcal{W}^B$ ($h_{IJ}(\phi)D_\mu \phi)^I(D_\mu \phi)^J$), and $U_{BC}$ and $V_{JK}$ are  unitary matrices defined as
\begin{equation}
\label{eq:UVmatrices}
U_{BC} = \left(
    \begin{array}{cccc}
    \frac{1}{\sqrt{2}} & \frac{1}{\sqrt{2}} & 0 & 0 \\
    \frac{i}{\sqrt{2}} & -\frac{i}{\sqrt{2}} & 0 & 0 \\
    0 & 0 & c_{\overline{\theta}} & s_{\overline{\theta}} \\
     0 & 0 & -s_{\overline{\theta}} & c_{\overline{\theta}} 
    \end{array}
    \right)\,,\,\,\,\,\,\,\,\,\,\,
V_{JK} = \left(
\begin{array}{cccc}
\frac{-i}{\sqrt{2}} & \frac{i}{\sqrt{2}} & 0 & 0\\
\frac{1}{\sqrt{2}} & \frac{1}{\sqrt{2}} & 0 & 0\\
0 & 0 & -1 & 0\\
0 & 0 & 0 & 1
\end{array}
\right)\,,
\end{equation}
with $c_{\overline{\theta}},s_{\overline{\theta}}$ also defined geometrically in terms of $\sqrt{g}^{AB}$  and $\alpha^A$ --- see \cite{Helset:2020yio} and Appendix \ref{sec:CONVENTIONS} for more details.  Finally the mass-eigenstate fields and couplings for the bosonic sector of the geo($\nu$)SMEFT are given by \cite{Helset:2018fgq,Helset:2020yio}
\begin{equation}
\label{eq:massbasisbosons}
   \mathcal{W}^{A,\mu} = \mathcal{U}_C^A \mathcal{A}^{C,\mu}\,, \,\,\,\,\,\,\,\,\,\, \phi^J = \mathcal{V}_K^J \Phi^K \,, \,\,\,\,\,\,\,\,\,\,\alpha^A = \mathcal{U}_C^A \beta^C\,,
\end{equation}
where $\mathcal{A}^{C,\mu}$ ($\Phi^K$) are the physical gauge (scalar) bosons of the SM,
\begin{equation}
    \mathcal{A}^{C,\mu} = \lbrace \mathcal{W}^+, \mathcal{W}^-, \mathcal{Z}, \mathcal{A} \rbrace\,, \,\,\,\,\,\,\,\,\,\, \Phi^{K} = \lbrace \Phi^-, \Phi^+,\chi, h \rbrace\,,
\end{equation}
and $\beta^C$ are the physical gauge couplings, 
\begin{equation}
\label{eq:masscouplings}
    \beta^C = \Biggl\{  \frac{g_2 \left(1-i\right)}{\sqrt{2}}, \frac{g_2\left(1+i\right)}{\sqrt{2}}, \sqrt{g_1^2 + g_2^2}\left(c^2_{\overline{\theta}} - s^2_{\overline{\theta}}\right), \frac{2 g_1 g_2}{\sqrt{g_1^2 + g_2^2}} \Biggl\}\,,
\end{equation}
written here with the couplings $g_{1,2}$ in the (unbarred) ($\nu$)SM limit, i.e. when higher-order operators do not contribute to their definition.  Their all-orders (barred) generalizations in the geometric framework are given in Appendix \ref{sec:CONVENTIONS} along with the geometric weak mixing angles appearing in \eqref{eq:UVmatrices} and \eqref{eq:masscouplings}.  In particular, these definitions --- written in terms of the all-orders field-space connections $g^{AB}$ and $h^{IJ}$ --- allow one to straightforwardly define the physical electroweak gauge boson masses (amongst other fundamental Lagrangian parameters) geometrically,
\begin{equation}
    \label{eq:allordergaugemass}
    \overline{m}_W^2 = \frac{\overline{g}_2^2}{4} \sqrt{h_{11}}^2 \overline{v}_T^2\,, \,\,\,\,\,\,\,\,\, \overline{m}_Z^2 = \frac{\overline{g}_Z^2}{4} \sqrt{h_{33}}^2 \overline{v}_T^2\,,\,\,\,\,\,\,\,\,\,\,\,\overline{m}_A^2 = 0\,,
\end{equation}
which can be expanded to arbitrary order in $\overline{v}_T/\Lambda$.

\subsection{Enumerating Two- and Three-Point Functions}
\label{sec:ENUMERATE}
As stated above we aim to enumerate the finite set of two- and three-point composite operator forms $f_i$ of the geo$\nu$SMEFT, where a `point' can a priori be a fermion $\psi$, a field-strength tensor, a Higgs-derivative term, or a fermion-derivative term:\footnote{Following \cite{Helset:2018fgq,Helset:2020yio}, the $\gamma_A$ matrices are electroweak symmetry generators written in a four-dimensional real representation, and their tilded notation (and that of the Levi-Civita tensors $\epsilon_{BC}$) implies that a gauge-coupling has been absorbed into the definition --- see Appendix \ref{sec:CONVENTIONS} for details.  $T^\mathcal{A}$ are the Gell-Mann matrices of QCD, $2T^{\pm}=\sigma_1 \pm i \sigma_2$ and $T_3=\sigma_3/2$ are isospin generators composed of Pauli matrices $\sigma_{\lbrace1,2,3\rbrace}$, and $Q_\psi = \sigma_3/2 + Y_\psi$ is the electric charge of the specified fermion in terms of its hypercharge $Y_\psi$, given in Table \ref{tab:nuSMEFTfieldandsymm}.}
\begin{align}
\nonumber
(D^\mu \phi)^I &= \left(\partial^\mu \delta^I_J -\frac{1}{2} \mathcal{W}^{A,\mu}\tilde{\gamma}^I_{A,J}\right) \phi^J \,, \\
\nonumber
D_\mu \psi &= \left[\partial_\mu + i \overline{g}_3 \mathcal{G}_{\mathcal{A}}^{\mu} T^\mathcal{A} + i \frac{\overline{g}_{2}}{\sqrt{2}}\left(\mathcal{W}^+ T^+ + \mathcal{W}^- T^-\right) + i \overline{g}_Z \left(T_3 - s_{\theta_Z}^2 Q_\psi\right)\mathcal{Z}_\mu + i Q_\psi \overline{e}\mathcal{A}^\mu \right] \psi \,,\\
\label{eq:pointsform}
\mathcal{W}^A_{\mu\nu} &= \partial_\mu \mathcal{W}^A_\nu -\partial_\nu \mathcal{W}^A_\mu - \tilde{\epsilon}^A_{BC} \mathcal{W}^B_\mu \mathcal{W}^C_\nu \,.
\end{align}
 Furthermore, we are only interested in identifying operators that are novel with respect to the geoSMEFT, and so we only need to identify functions whose $f_i$ have an explicit dependence on the sterile gauge-singlet $N$.  Finally, given that $N$ is a Lorentz spinor, the relevant $f_i$ must come with at least two fermion-dependent points, since $G_i$ only has scalar field dependence.

Considering these simple constraints, one can rapidly enumerate the two- and three-point composite operators $f_i$ that fulfill them, finding 
\begin{itemize}
\item a Yukawa operator of the form {${\mathcal{Y}_N(\phi)\,\overline{N}\,\ell}$},
\item a Majorana mass operator of the form {${\eta_N(\phi)\,  \overline{N}\, N^c}$},
\item dipole-type operators of the form {${d_{\psi_1\psi_2}(\phi)\, \psi_1\, \sigma_{\mu \nu} \,\psi_2 \,\mathcal{W}^{\mu\nu}}$} with $\psi_1\psi_2 \in \lbrace \overline{N}\ell, \overline{e}N^c, \overline{N} N^c \rbrace$,
\item single-derivative operators of the form {${ L_{\psi N}(\phi)\, \left(D^\mu \phi \right) \psi_1 \,\gamma_\mu \,\psi_2}$} with $\psi_1 \psi_2 \in \lbrace \overline{e}N, \overline{N}N,\ell \mathbb{C} N \rbrace$,
\end{itemize}
as well as hermitian-conjugate combinations of said fields, when relevant.  We now study the saturation of the $f_i$ listed above by utilizing techniques embedded in the Hilbert Series (HS), and the novel (automated) HS generator {\tt{ECO}} \cite{Marinissen:2020jmb}.  {\tt{ECO}} is a {\tt{FORM}} \cite{Vermaseren:2000nd,Ruijl:2017dtg} program that exploits the Molien-Weyl formula for computing HS, and which speeds up the computation by orders of magnitude in comparison to prior similar approaches (see e.g. \cite{Henning:2015daa,Henning:2015alf,Henning:2017fpj}).  It has built-in support for counting EFT operators with (B)SM particle content and SM gauge symmetries, as well as additional gauge or global U(1) symmetries as defined by the user.  For example, we have computed the {\tt{ECO}} counting of $\nu$SMEFT operators up to $d=11$ and $n_f = n_l = 3$ in Table \ref{tab:nuSMEFTcounting}.\footnote{Whilst trivial, we are unaware of any counting of $\nu$SMEFT operators up to this mass dimension present in the literature.  Note that generating the $n_f = 3$ counting of 11347838 $d=11$ $\nu$SMEFT operators required only 1.57 seconds of computing time on a standard laptop (!)}

\begin{table}
\centering
{\renewcommand{\arraystretch}{1.4}
\begin{tabular}{|c||c|c|c|c|c|}
\hline
\multicolumn{6}{|c|}{geo$\nu$SMEFT Composite Operator Saturation} \\
\hline
Mass Dimension &  $d_0$ & $d_0+2$ & $d_0+4$ & $d_0+6$ & $d_0+8$ \\
\hline
\hline
$\mathcal{Y}_N(\phi)\, \overline{N} \ell$ + h.c. & $2\, n_f\cdot n_l$ & $2\, n_f\cdot n_l$ & $2 \,n_f\cdot n_l$ & $2 \,n_f\cdot n_l$ &  $2 \,n_f\cdot n_l$ \\
\hline
$d_{N\ell}(\phi)\,\overline{N}\, \sigma_{\mu\nu}\,\ell \,\mathcal{W}^{\mu\nu}$ + h.c. & $4 \,n_f \cdot n_l$ & $6\, n_f \cdot n_l$ & $6\, n_f \cdot n_l$ & $6\, n_f \cdot n_l$ & $6\, n_f \cdot n_l$ \\
\hline
$L_{eN}(\phi)\,\,(D^\mu \phi)\,\overline{e}\,\gamma_\mu\, N$ + h.c. & $2\,n_f \cdot n_l$ & $2\,n_f \cdot n_l$ & $2\,n_f \cdot n_l$ & $2\,n_f \cdot n_l$ & $2\,n_f \cdot n_l$ \\
\hline
$L_{NN}(\phi)\,(D^\mu \phi)\,\overline{N} \gamma_\mu N$ & $n_f^2$  & $n_f^2$  & $n_f^2$ & $n_f^2$  & $n_f^2$ \\
\hline
\hline
$\eta_N (\phi)\, \overline{N} N^c$ + h.c. & $(n_f+n_f^2)$ & $(n_f+n_f^2)$ & $(n_f+n_f^2)$ & $(n_f+n_f^2)$ & $(n_f+n_f^2)$ \\
\hline
$d_{eN}(\phi)\,\overline{e}\, \sigma_{\mu\nu}\, N^c \,\mathcal{W}^{\mu\nu}$ + h.c. & $2 \,n_f \cdot n_l$ & $2\, n_f \cdot n_l$ & $2\, n_f \cdot n_l$ & $2\, n_f \cdot n_l$ & $2\, n_f \cdot n_l$ \\
\hline
$ d_{NN}(\phi)\, \overline{N} \sigma_{\mu \nu} N^c \, \mathcal{W}^{\mu\nu}$ + h.c. & $\frac{1}{2} (n_f+n_f^2)$ &$(n_f+n_f^2)$ & $(n_f+n_f^2)$ & $(n_f+n_f^2)$ & $(n_f+n_f^2)$ \\
\hline
$L_{\ell N}(\phi)\,\,(D^\mu \phi)\,\ell\mathbb{C}\,\gamma_\mu\, N$ + h.c. & $4\,n_f \cdot n_l$ & $4\,n_f \cdot n_l$ & $4\,n_f \cdot n_l$ & $4\,n_f \cdot n_l$ & $4\,n_f \cdot n_l$ \\
\hline
\end{tabular}}
\caption{
Saturation of composite geo$\nu$SMEFT operators in mass dimension, for arbitrary numbers of sterile neutrino flavors $n_f$ and SU(2)$_L$ lepton doublet flavors $n_l$.  For the Yukawa and LNV mass operators the relevant starting dimension is $d_0=4$ (including the Majorana mass matrix), while the dipole and derivative-type operators turn on at varying mass dimensions: $d_0(d_{NN}) = 5$, $d_0(d_{N\ell},L_{eN},L_{NN}) = 6$, $d_0(d_{eN},L_{\ell N}) =7$. The table is organized into even- and odd-dimensional $f_i$.
\label{tab:nuSMEFTsaturation}
}
\end{table}
If the field-space connections associated to the two- and three-point operators enumerated above are truly defined at all-orders in $\overline{v}_T/\Lambda$, one must demonstrate that the number of independent operators (accounting for all flavor and gauge indices) constituting the composite operator forms $f_i$ \emph{saturate} at a finite value in mass dimension.  We have shown that this is indeed the case for all of the functions defined in Section \ref{sec:ENUMERATE}, with the results presented in Table \ref{tab:nuSMEFTsaturation} for mass dimensions up to $d_0 + 8$, where $d_0$ is the mass dimension where the composite operator first appears in the $\nu$SMEFT.  Note that in what follows we begin the counting of the operators contributing to a given connection at the first order they appear in the non-renormalizable OPE, but in Section \ref{sec:CONNECTIONS} we will include any renormalizable contributions as well, for completeness.
\subsection{Saturation in $\overline{v}_T/\Lambda$:  Even-Dimensional Operators}
\label{sec:SATURATE}
\subsubsection*{{\bm{$\mathcal{Y}_N(\phi)\,\overline{N}\,\ell$}}}
One observes that, as expected, the Yukawa operators of the form
\begin{equation}
    \left[\mathcal{Q}_{NH}^{(6+2n)}\right]_{pr} = \left(H^\dagger H\right)^{n+1}\,\tilde{H}^\dagger \left(\overline{N}_p \ell_r \right)
\end{equation}
saturate immediately in the $1/\Lambda$ expansion as a function of the number of independent coefficients in the $p \times r$ flavor matrix:  $n_f\cdot n_l$ ($\times 2$ to account for the Hermitian conjugate matrix $\propto \overline{\ell}N$).
\subsubsection*{{\bm{$d_{N\ell}(\phi)\,\overline{N}\, \sigma_{\mu\nu}\, \ell \, \mathcal{W}^{\mu\nu}$}}} 
At mass dimension six there are contributions to the first dipole operator from both $B$ and $W^a$ couplings,
\begin{align}
    \left[\mathcal{Q}^{(6+2n)}_{N\ell W}\right]_{pr} &= i \left(H^\dagger H\right)^n \tilde{H}^\dagger \sigma^A \left(\overline{N}_p\, \sigma_{\mu\nu} \,\ell_r\, \mathcal{W}_A^{\mu\nu}\right) \,,
\end{align}
while, beginning at mass dimension 8, additional couplings to the SU(2)$_L$ bosons are allowed via operators of the form
 \begin{align}
    \left[\mathcal{Q}^{(8+2n)}_{N\ell W2}\right]_{pr} &= -i \left(H^\dagger H\right)^n \left(\tilde{H}^\dagger \sigma^A H\right) H^\dagger \left(\overline{N}_p\, \sigma_{\mu\nu} \, \ell_r\, \mathcal{W}_A^{\mu\nu}\right)\left(1-\delta_{A4}\right) \,.
\end{align}
 That two additional sets of composite operator forms, accounting for novel couplings to $W^a$ (but with different SU(2)$_L$ contractions), enter at dimension eight and beyond explains the jump in counting in Table \ref{tab:nuSMEFTsaturation} between $d_0$ and $d_0 + 2$ by a factor of $3/2$. For example, for $n_f = n_l = 3$, the $W_a$ boson couplings contribute 18 operators (including Hermitian conjugate structures) at $d = 6$ and 36 operators at $d = 8$ and beyond, while the $B$-field couplings contribute 18 operators at all even mass dimensions starting at $d=6$.
\subsubsection*{{\bm{$L_{eN}(\phi)\,\left(D^\mu \phi \right) \overline{e} \, \gamma_\mu \,N$}}}
The operator saturation for this derivative operator occurs immediately in mass dimension, as the sole contributors are from operators of the form
\begin{equation}
\left[\mathcal{Q}^{(6+2n)}_{DeN}\right]_{pr} = -\left(H^\dagger H \right)^n\left(H^\dagger\,iD^\mu\,\tilde{H}\right)\left(\overline{e}_p\,\gamma_\mu\,N_r\right)
\end{equation}
and their hermitian conjugates, which contribute at mass dimension six and above.
\subsubsection*{{\bm{$ L_{NN}(\phi) \left(D^\mu \phi \right) \overline{N} \gamma_\mu N$}}}
The saturation for this set of derivative operators is again immediate in mass dimension, following from $\nu$SMEFT operators of the form
\begin{equation}
    \left[\mathcal{Q}^{(6+2n)}_{DNN}\right]_{pr} = \left(H^\dagger H\right)^n \left(H^\dagger\, i D^\mu \,H \right)\left(\overline{N}_p \gamma_\mu N_r \right)\,,
\end{equation}
which generate $n_f^2$ contributions.
\subsection{Saturation in $\overline{v}_T/\Lambda$:  Odd-Dimensional Operators}
\subsubsection*{{\bm{$\eta_N(\phi)\, \overline{N}\, N^c$}}}
The LNV connection $\eta_N(\phi)$ is a complex symmetric matrix in flavor space, and so there are fewer degrees of freedom (operators) of the form
\begin{equation}
    \left[\mathcal{Q}_{NN}^{(5+2n)}\right]_{pr} = \left(H^\dagger H\right)^{n+1} \left(\overline{N}_p N^c_r\right)
\end{equation}
in comparison to Yukawas, namely $n_f(1+n_f)/2$ (again $\times 2$ to account for operators $\propto \overline{N^c} N$), although, as with the Yukawas, operator saturation occurs immediately in mass dimension.
\subsubsection*{{\bm{$d_{eN}(\phi)\,\overline{e}\, \sigma_{\mu\nu}\, N^c \, \mathcal{W}^{\mu\nu}$}}} 
The operators contributing to this RH connection are of the form
\begin{align}
    \left[\mathcal{Q}^{(7+2n)}_{eNW}\right]_{pr} &= i\left(H^\dagger H\right)^n \left(H^\dagger \sigma^A \tilde{H}\right)\left(\overline{e}_p\, \sigma_{\mu\nu} \, N^c_r\, \mathcal{W}_A^{\mu\nu}\right)\left(1-\delta_{A4}\right) \,,
\end{align}
where it is clear that they begin at mass-dimension seven, and that there are no couplings to the hypercharge gauge boson $B$ at any mass dimension. Saturation in $1/\Lambda$ is immediate.
\subsubsection*{{\bm{$d_{NN}(\phi)\, \overline{N}\,\sigma_{\mu\nu}\, N^c\,\mathcal{W}^{\mu\nu}$}}}
This set of dipole terms saturate in a way that is analogous to $d_{N\ell}$ above, in that an additional class of operators turns on at next-to-leading order in the $1/\Lambda$ OPE.  To see this, note that at dimension five the only contributions to $f_i$ come from operators of the form 
\begin{align}
    \left[\mathcal{Q}^{(5+2n)}_{NNB}\right]_{pr} &= i\left(H^\dagger H\right)^n \sigma^A\left(\overline{N}_p \,\sigma_{\mu\nu}\, N^c_r\, \mathcal{W}_A^{\mu\nu}\right)\,\delta_{A4} \,,
\end{align}
which only couple to the U(1)$_Y$ gauge boson $B$, whereas beginning at dimension seven one also finds contributions to $f_i$ from
 \begin{align}
    \left[\mathcal{Q}^{(7+2n)}_{NNW}\right]_{pr} &= i\left(H^\dagger H\right)^n \left(H^\dagger \sigma^A H\right)\left(\overline{N}_p \,\sigma_{\mu\nu}\, N^c_r\, \mathcal{W}_A^{\mu\nu}\right)\left(1-\delta_{A4}\right) \,,
\end{align}
i.e. from the SU(2)$_L$ gauge bosons $W_a$.  Hence one sees the number of contributing operators jump by a factor of two in Table \ref{tab:nuSMEFTsaturation} for this sector.  However, after dimension 7 there are no further contributions, and the counting saturates at $(n_f + n_f^2)$ as expected.
\subsubsection*{{\bm{$L_{\ell N}(\phi)\,\left(D^\mu \phi / \tilde{H}^\dagger \right) \ell \mathbb{C}\, \gamma_\mu \, N$}}}
This derivative class is unique in that, in the basis of \cite{Li:2021tsq}, the connection $L_{\ell N}$ can be understood to contain two independent sub-classes of composite operators $f_i$ in the geometric formulation, $L_{\ell N}(\phi) \supset \lbrace L_{\ell N1}(\phi), L_{\ell  N2}(\phi)\rbrace$, for which two independent field-space connections can be written down in Section \ref{sec:CONNECTIONS}.  The operator classes are given by
\begin{align}
\label{eq:LLN1}
\left[\mathcal{Q}^{(7+2n)}_{D\ell N1}\right]_{pr} &= \left(H^\dagger H \right)^{n+1}\left(iD^\mu\,\tilde{H}^\dagger\right)\left(\ell_p \mathbb{C}\,\gamma_\mu\,N_r\right)\,, \\
\label{eq:LLN2}
\left[\mathcal{Q}^{(7+2n)}_{D\ell N2}\right]_{pr} &= \left(H^\dagger H \right)^{n}\left(H^\dagger \,i D^\mu \, H\right)\tilde{H}^\dagger\left(\ell_p \mathbb{C}\,\gamma_\mu\,N_r\right)\,,
\end{align}
where the differentiating feature is the SU(2)$_L$ contractions of the Higgs-derivative terms, which occurs through the SU(2)$_L$ lepton doublet $\ell$ in \eqref{eq:LLN1}, as opposed to the additional conjugate Higgs field $\tilde{H}^\dagger$ in \eqref{eq:LLN2}.  The latter contraction for $L_{\ell  N2}(\phi)$ generates an $SU(2)_L$ scalar with polynomial dependence on the real coordinates $\phi_I$ from the outset of the geometric factorization, leaving an explicit $D^\mu \phi$ term in the $f_i$, whereas our geo$\nu$SMEFT notation leaves the real Higgs-coordinate dependence implicit in the $D^\mu \tilde{H}^\dagger$ term appearing in the $f_i$ for \eqref{eq:LLN1}.\footnote{This is analogous to the appearance of $\tilde{H}^\dagger(\phi)$ in \eqref{eq:YukawaConnection} below, for the Yukawa connection $\mathcal{Y}_N(\phi)$.}   Regardless, both operator classes in \eqref{eq:LLN1}-\eqref{eq:LLN2} individually contribute $2\, n_f \cdot n_l$ terms in the saturation presented in Table \ref{tab:nuSMEFTsaturation}, (totaling $4\, n_f \cdot n_l$ terms), which occurs immediately in the $\overline{v}_T/\Lambda$ expansion.

\section{geo$\nu$SMEFT Field-Space Connections}
\label{sec:CONNECTIONS}
Having enumerated the two- and three-point operator classes with well-defined connections in field space, and shown that they saturate in mass-dimension as they must to be defined at all orders in the $\overline{v}_T/\Lambda$ expansion, we can now write down the exact functional forms of their associated field space connections.  In what follows, we do so for each operator found in Section \ref{sec:ENUMERATE}, which can generically be defined as a variation of the SMEFT Lagrangian with respect to the field combinations appearing in the $f_i$ under question, 
\begin{equation}
\label{eq:connectvariation}
    G_i \equiv \frac{\delta^k  \mathcal{L}_{(\nu)\text{SMEFT}}}{\prod_{i=1}^{\hat{k}} \delta \hat{f
}_i} \Big \vert_{\mathcal{L}(\alpha,\beta,..) \rightarrow 0}\,,
\end{equation}
where $\hat{k},\hat{f}$ indicates a  modified `pointiness' (counting fermion bilinears as a single variation) of the composite operator, and where the $\mathcal{L}(\alpha,\beta,..) \rightarrow 0$ notation implies that all spin connections and Lagrangian terms, dependent on effective gauge couplings ($\alpha, \beta, ...$) and carrying non-trivial Lorentz indices, are sent to zero.  For example, the Higgs potential connection in the geo($\nu$)SMEFT is trivially given by $V(\phi) = - \mathcal{L}_{\text{SMEFT}} \vert_{\mathcal{L}(\alpha, \beta, ...)\rightarrow 0}$, while the $\mathcal{Y}_N(\phi)$, $d_{N\ell}(\phi)$, and $L_{eN}(\phi)$ connections from above are respectively defined as
\begin{equation}
\label{eq:samplevariations}
    \Biggl\{ \frac{\delta \mathcal{L}_{\nu\text{SMEFT}}}{\delta \left(\overline{N}_{p}\ell_{r}\right)},\,\,\,
    \frac{\delta^2 \mathcal{L}_{\nu\text{SMEFT}}}{\delta (\overline{N}_p \sigma_{\mu\nu}\ell_r)\,\delta\mathcal{W}^{\mu\nu}},\,\,\,
    \frac{\delta^2 \mathcal{L}_{\nu\text{SMEFT}}}{\delta (D^\mu \phi)\delta(\overline{e}_p \gamma_\mu N_r)} 
    \Biggl\} \Biggl \vert_{\mathcal{L}(\alpha,\beta,..) \rightarrow 0}\,\,\,,
\end{equation}
where again $\lbrace p,r \rbrace$ are flavor labels.  Hermitian conjugate connections can of course also be similarly defined when relevant.  In what follows we will not explicitly list all of the variational definitions analogous to \eqref{eq:samplevariations}, but instead give the resulting analytic form for each connection.

Note that in doing so one can work out the following conversions between Higgs field contractions, when moving to the real $\phi_i$ coordinate basis:
\begin{align}
\nonumber
H^\dagger H &= \frac{1}{2} \phi_I \phi^I \equiv \frac{1}{2} \phi^2\,, \,\,\,\,\,\,\,\,\,\,\,\,\,\,\,\,\,\,\,\,\,\,\,\,\,\,\,\,\,\,\,\,\,\,\,\,\,\,\,\,\,\,\, H^\dagger \,\sigma^A\,H = -\frac{1}{2} \phi_I\,\Gamma_{A,J}^I\,\phi^J\,, \\
\nonumber
\tilde{H}^\dagger\,\sigma^A\,H &=\frac{1}{2} \tilde{\phi}_I\left(-\Gamma_{A,J}^I + i \gamma_{A,J}^I \right) \phi^J\,, \,\,\,\,\,\,\,\,\,\,\,\,\,\,\, H^\dagger\,iD^\mu\,H = -\frac{1}{2} \phi_I \left(i \Gamma_{4,J}^I + \gamma_{4,J}^I \right)(D^\mu \phi)^J \,, \\
 \label{eq:realconversion}
H^\dagger\,\sigma^A\,\tilde{H} &= \frac{1}{2} \tilde{\phi}_I\left(-\Gamma_{A,J}^I - i \gamma_{A,J}^I \right) \phi^J \,, \,\,\,\,\,\,\,\,\,\,\,\,\,\,\,
H^\dagger\,iD^\mu\,\tilde{H} = -\frac{1}{2} \phi_I \left(i \Gamma_{4,J}^I + \gamma_{4,J}^I \right)(D^\mu \tilde{\phi})^J 
\end{align}
Here $\tilde{\phi} = \lbrace \phi_3, \phi_4, -\phi_1, -\phi_2 \rbrace$.  Given \eqref{eq:connectvariation} and \eqref{eq:realconversion} one can then quickly derive the all-orders field-space connections for the operator classes identified in Section \ref{sec:OPERATORS}, which we do below for each $f_i$.  Note that we absorb any dependence on the implied new physics scale $\Lambda$ into the definition of the Wilson coefficients, as in (e.g.) \eqref{eq:SMEFTwithWeinberg}.  We also organize the following presentation based on the class of physics operators found in Section \ref{sec:OPERATORS} (i.e. mass-type, dipole-type, or derivative-type), as opposed to by the (odd or even) mass dimension that they enter the $\nu$SMEFT OPE.
\subsection{Mass-Type Operators}
\label{sec:masstype}
The Yukawa-like field-space connection $\mathcal{Y}_N$ defined above is given by
\begin{equation}
\label{eq:YukawaConnection}
    \mathcal{Y}_N(\phi)_{pr} = - \tilde{H}^\dagger(\phi_I) \left[Y_N\right]_{pr}^\dagger +  \tilde{H}^\dagger(\phi_I) \sum_{n=0}^{\infty}\, \tilde{C}_{\underset{pr}{NH}}^{(6+2n)}\,\left(\frac{\phi^2}{2} \right)^{n+1}
\end{equation}
in its closed-form, all-orders expression.  One observes that the leading contribution to $\mathcal{Y}_N(\phi)$ is from the $\nu$SM Yukawa operator given in \eqref{eq:LagSS}, and that $\tilde{H}^\dagger(\phi)$ remains in its (complex) two-component doublet form (as opposed to a four-component real vector $\phi_I$), since it must still be contracted with the SU(2)$_L$ doublet $\ell$ in the composite operator form factored out of this expression in the geo($\nu$)SMEFT formalism.  On the other hand, the infinite tower of $H^\dagger H$ dressings has been contracted and converted to the real coordinates of the Higgs field space. 
\\
\\
Similarly, the LNV Majorana mass field-space connection $\eta_N(\phi)$ is given by
\begin{equation}
\label{eq:MajoranaMassConnection}
    \eta_N(\phi)_{pr} = -\frac{1}{2} \left[M_N\right]_{pr} + \sum_{n=0}^{\infty} \,\tilde{C}_{\underset{pr}{NN}}^{(5+2n)}\, \left(\frac{\phi^2}{2}\right)^{n+1}
\end{equation}
where we have kept the factor of $1/2$, due to the Majorana nature of the mass matrix appearing at leading order, in the renormalizable term. 
\subsection{Dipole-Type Operators} 
There are three dipole-like composite operator forms appearing in Table \ref{tab:nuSMEFTsaturation}, with the first a coupling between  $\lbrace \overline{N}, \ell, \mathcal{W}^{\mu\nu} \rbrace$, the second between $\lbrace \overline{e}, N^c, \mathcal{W}^{\mu\nu} \rbrace$, and the third between $\lbrace \overline{N}, N^c, \mathcal{W}^{\mu\nu} \rbrace$.  The associated field-space connections are respectively given by
\begin{align}
\nonumber
    d_{N\ell} (\phi)_{pr} &=i \sum_{n=0}^\infty \left[\tilde{H}^\dagger(\phi)\,\sigma^A\, \,\tilde{C}_{\underset{pr}{N \ell W}}^{(6+2n)} +\frac{\tilde{\phi}_I}{2} \left(\Gamma_{A,J}^I - i \, \gamma_{A,J}^I \right) \phi^J \,\left(1-\delta_{A4} \right) H^\dagger(\phi)\,\tilde{C}_{\underset{pr}{N\ell W2}}^{(8+2n)} \right]\left(\frac{\phi^2}{2}\right)^n \,,
    \end{align}
for the even connection, and
\begin{align}
\nonumber
    d_{eN}(\phi)_{pr} &= i\sum_{n=0}^\infty\left[\frac{\tilde{\phi}_I}{2}\, \left(\Gamma_{A,J}^I + i \gamma_{A,J}^I\right)\,\phi^J \,\left(1-\delta_{A4}\right)\,\tilde{C}_{\underset{pr}{eN W}}^{(7+2n)}\right]\left(\frac{\phi^2}{2}\right)^n\,, \\
\label{eq:dipoleconnections}
    d_{NN}(\phi)_{pr} &= i\sum_{n=0}^{\infty} \left[ \sigma^A\,\delta_{A4}\,\tilde{C}_{\underset{pr}{NNB}}^{(5+2n)} - \frac{\phi_I}{2}\,\Gamma^{I}_{A,J}\,\phi^J\left(1-\delta_{A4}\right) \,\tilde{C}_{\underset{pr}{NNW}}^{(7+2n)}\right]\left(\frac{\phi^2}{2}\right)^n\,,
\end{align}
for the odd connections.  Unlike the Yukawa and Majorana connections of Section \ref{sec:masstype}, it's clear that these dipole-like connections have support only at the non-renormalizable level, with the earliest contribution coming at $d=5$ for the RH connection $d_{NN}(\phi)$.  One also clearly observes the structure associated to two different operator types contributing to the saturation of $d_{NN}(\phi)$ and $d_{N\ell}(\phi)$, discussed in Section \ref{sec:OPERATORS} and visible in Table \ref{tab:nuSMEFTsaturation}. 
\subsection{Derivative-Type Operators}
Distinguishing both sub-classes of operators associated to $L_{\ell N}(\phi)$, there are four additional composite operator forms $f_i$ with a derivative dependence appearing in Table \ref{tab:nuSMEFTsaturation}. They couple $\lbrace D^\mu \phi, \overline{e}, N \rbrace$, $\lbrace D^\mu \phi, \overline{N}, N\rbrace$, $\lbrace D^\mu\phi, \ell \mathbb{C}, N \rbrace$, and $\lbrace  D^\mu \tilde{H}^\dagger, \ell \mathbb{C}, N \rbrace$, and their respective field-space connections are given by  
\begin{align}
   \nonumber
   L_{eN}(\phi)_{pr} &= \sum_{n=0}^\infty\left[\frac{\phi_I }{2} \left(i\, \Gamma_{4,J}^I + \gamma_{4,J}^I \right)\,\,\tilde{C}_{\underset{pr}{DeN}}^{(6+2n)} \right]\left(\frac{\phi^2}{2}\right)^n\,,\\
 \nonumber
 L_{NN}(\phi)_{pr} &= -\sum_{n=0}^{\infty}\left[ \frac{\phi_I}{2} \left(i\, \Gamma^I_{4,J}+\gamma^I_{4,J}\right)  \,\tilde{C}_{\underset{pr}{DNN}}^{(6+2n)}\, \right]\left(\frac{\phi^2}{2} \right)^{n}\,,
 \end{align}
 for the even connections and
 \begin{align}
  \nonumber
   L_{\ell N1}(\phi)_{pr} &= \sum_{n=0}^\infty \tilde{C}_{\underset{pr}{D\ell N1}}^{(7+2n)}  \left(\frac{\phi^2}{2}\right)^{n+1}\\ 
\label{eq:derivative connections}
   L_{\ell N2}(\phi)_{pr} &= -\sum_{n=0}^\infty\left[ \frac{\phi_I}{2} \left(i\, \Gamma^I_{4,J}+\gamma^I_{4,J}\right)\,\tilde{H}^\dagger \,\tilde{C}_{\underset{pr}{D\ell N2}}^{(7+2n)}\right]\left(\frac{\phi^2}{2}\right)^n\,,
\end{align}
for the odd connections, which are again all fundamentally non-renormalizable objects whose support begins only at $d=6$ or $d=7$ in the $1/\Lambda$ expansion of the $\nu$SMEFT.
\subsection*{Summary Comments}
We have presented a complete list of field-space connections $G_i$ associated to the novel ($N$-dependent) two- and three-point interactions introduced in the $\nu$SMEFT.  These objects are defined at \emph{all-orders} in the $\overline{v}_T/\Lambda$ expansion of the $\nu$SMEFT's Lagrangian \eqref{eq:nuSMEFT}, and can be (if desired) trivially expanded to any fixed-order in said expansion as may be required for phenomenology.  Upon $H(\phi)$ acquiring its vev, the $G_i$ reduce to a number and emissions of the physical Higgs field $h$.  Some additional comments are in order:
\begin{itemize}
    \item The $G_i$ are defined on the field spaces formed by the composite operators presented in Section \ref{sec:OPERATORS}, and hence these new objects will contribute to the overall field-space geometry of the geo$\nu$SMEFT.  However, we have not made an attempt to formally define any novel geometric structures (e.g. metrics, Christoffel symbols, Riemmann curvature tensors, covariant derivatives..) that may be associated to them, beyond those implied in \cite{Helset:2020yio}, as doing so in fermionic theories represents an active line of research.  It is true that, as in \cite{Helset:2020yio}, sending $C \rightarrow 0$ in the geo($\nu$)SMEFT amounts to a projection to the renormalizable, `flat' ($\nu$)SM of \eqref{eq:nuSMEFT}, where at least this notion of `flatness' can be defined with respect to curvature tensors derived from the same (bosonic) metrics $h_{IJ}$ and $g_{AB}$ that appear in the geoSMEFT \cite{Helset:2020yio}, or perhaps even the unified gauge-Higgs metric recently explored in \cite{Helset:2022tlf,Helset:2022pde}. Then (field-redefinition-invariant) physical scattering amplitudes become simple functions of the descendant geometric objects (e.g. curvature tensors and covariant derivatives) derived from said metric(s). In this setup, at least the non-derivative fermionic `mass-type' $G_i$ defined above might be thought of as contributions to a generalized all-$\overline{v}_T/\Lambda$-orders potential term, cf. \cite{Helset:2022tlf,Finn:2020nvn}, which will also contribute to scattering amplitudes and RG-evolved geometric objects --- see \cite{Helset:2022tlf} and the comments below.  On the other hand, defining explicit fermionic field-space metric(s) analogous to (or in unison with) the scalar and gauge sectors is potentially complicated by the fact that fermions have first-order equations of motion and are also non-commutative, Grassmannian fields. The authors of \cite{Finn:2020nvn} have recently presented a formalism based on supermanifolds \cite{Batchelor:1978fz,Leites:1980rna} that they argue circumvents these issues, yielding well-defined metrics and descendant geometric quantities, and ultimately a manifestly reparameterization-invariant theory. Extending this type of analysis to the all-$\overline{v}_T/\Lambda$-orders geo($\nu$)SMEFT would be interesting future work. Another option may be to extend the kinematics-dependent geometric formalism of \cite{Cheung:2022vnd} to the geo($\nu$)SMEFT. Generally speaking though, a more robust understanding of the mathematical properties of fermionic $G_i$, how to understand their role in implied field-space geometries, and how to exploit them phenomenologically represent active lines of research in the literature.  Regardless, the factorization and definitions presented in Sections \ref{sec:OPERATORS}-\ref{sec:CONNECTIONS} above hold independently of any such embedding.
    \item Given their all-orders definition, the $G_i$ can be understood to absorb all the (otherwise arbitrary) Wilson coefficients $C$ present into a single object, thereby effectively reducing the number of free parameters in the EFT, at least at tree level.  However, realistic phenomenology requires an understanding of the RG structure of an EFT, as is known for (e.g.) the $d=6$ SMEFT \cite{Jenkins:2013zja,Jenkins:2013wua,Alonso:2013hga}.  How the $G_i$ defined above and in \cite{Helset:2020yio} behave under RG evolution --- and critically whether or not an all-$\overline{v}_T/\Lambda$-orders structure is preserved under said RGE --- remains an open question in the literature for arbitrary $G_i$.  However, the RGE for (e.g.) Riemman curvature tensors formed from bosonic field-space metrics has been recently computed in \cite{Helset:2022tlf,Helset:2022pde} for the unified scalar-gauge effective theory defined therein, revealing that  the geometric objects (including an auxiliary potential) appearing at tree level also control the theory's scale evolution.  In this sense the RGE generates a deformation of the implied tree-level geometry, and this should also be true for the connections  $G_i$ (and any derivative objects) defined at all-$\overline{v}_T/\Lambda$-orders as above. Demonstrating this will establish a Lagrangian and subsequent scattering amplitude predictions that hold at all-$\overline{v}_T/\Lambda$-orders \emph{and} all scales, and again represents future work.
    \item While we have only discussed the finite list of two- and three-point operators, connections can be readily derived for higher-point functions as well.  This means that, for a particular physical process, one can straightforwardly obtain an all-$\overline{v}_T/\Lambda$-orders, tree-level amplitude in the geo($\nu$)SMEFT.
    \item While we have only discussed electroweak interactions with the gauge singlet, strong interactions are also possible, albeit in higher-point $f_i$ --- see e.g. \cite{Li:2021tsq}.
    \item In \eqref{eq:pointsform} we mentioned that operators with fermion-derivative terms can in principle also contribute to the two- and three-point geo($\nu$)SMEFT, though no such objects were found in Section \ref{sec:OPERATORS}. Yet fermion-derivative term(s) appear in the $d=7$ $\nu$SMEFT bases presented in \cite{Bhattacharya:2015vja,Liao:2016qyd}, not to mention the $d=7$ SMEFT bases presented in \cite{Lehman:2014jma,Liao:2016hru}. However, one can manipulate derivative operators in two- and three-point interactions such that they only strike Higgs fields, and in a manner where successive insertions of derivatives on operators with two fermions (e.g. $\mathcal{O} \sim D^{2n} H^2 \psi^2$) only contribute to $(n >3)$-point interactions --- see \cite{Li:2021tsq} (\cite{Li:2020xlh}) for evidence of this in the $\nu$SMEFT (SMEFT).\footnote{I thank Adam Martin for this important comment and insight. Also see the dim $= 9$ SMEFT basis in \cite{Liao:2020jmn}.} Note that this also has implications for the all-$\overline{v}_T/\Lambda$-orders structure of field-space metrics derived for fermionic theories using (e.g.) the supermanifold methods of \cite{Finn:2020nvn} mentioned above.
    \item To our knowledge, $\eta_N$, $d_{(eN,NN)}$, and $L_{\ell N}$  represent the first odd-dimensional field-space connections derived in the literature.  It would be interesting to derive analogous forms in the geoSMEFT, and pursue their phenomenology as well. 
\end{itemize}
The final commentary regarding the derivation of all-order Feynman rules in specific bases will be left to Section \ref{sec:PHENO} below.

\section{Towards Phenomenology at All $\overline{v}_T/\Lambda$ Orders}
\label{sec:PHENO}
While the results of Sections \ref{sec:OPERATORS}-\ref{sec:CONNECTIONS} represent the core output of this study, in this Section we briefly outline the route to calculating all-$\overline{v}_T/\Lambda$-orders amplitudes in the geo$\nu$SMEFT defined above, which will be pursued in more detail in later work.

Indeed, besides the benefits mentioned at the end of Section \ref{sec:CONNECTIONS}, the geometric formulation presented above also allows for the immediate derivation of Feynman rules which are themselves  valid at all orders in $\overline{v}_T/\Lambda$. 
For example, upon gauge-fixing the geo($\nu$)SMEFT with the Background Field Method (BFM) \cite{DeWitt:1967ub,tHooft:1973bhk,Abbott:1981ke,Shore:1981mj,Denner:1994xt},\footnote{See e.g. \cite{Hartmann:2015oia,Dekens:2019ept,Buchalla:2019wsc,Corbett:2020bqv} for prior studies utilizing the BFM in SMEFT contexts, and \cite{Corbett:2019cwl,Corbett:2020bqv,Corbett:2020ymv} where it is shown that the BFM allows for the preservation of Ward identities in the SMEFT. I thank Tyler Corbett for his help understanding Feynman rules in geometric EFTs.} which effectively doubles the bosonic field content of the Lagrangian,
\begin{equation}
\label{eq:BFM}
    \mathcal{L}_{\text{cl}}\left(\phi^I, \mathcal{W}^A, \mathcal{G}^A, \psi\right) \,\,\,\longrightarrow \,\,\, \mathcal{L}_{\text{cl}}\left(\phi^I + \hat{\phi}^I, \mathcal{W}^A + \hat{\mathcal{W}}^A, \mathcal{G}^A + \hat{\mathcal{G}}^A, \psi\right)\,,
\end{equation}
one is capable of computing amplitudes whose external particles are all classical background fields (the hatted quantities in \eqref{eq:BFM}), while internal propagators are quantum fields.  See (e.g.) \cite{Corbett:2021jox} for a complete one-loop calculation with the BFM in the geoSMEFT.

Critically, in the geometric formalism, the field-space connections amount to all-order vertices whilst the composite operator forms source any non-trivial momentum dependence that may appear.  In order to extract the former, all one needs to do is take the variation of the connection with respect to the physical (external) higgs field $\hat{h}$, and then the expectation value of the resulting object:
\begin{equation}
\mathbb{F}(\mathcal{O}(G_i\,f_i)) \propto \biggl< \frac{\delta G_i}{\delta\hat{h}}\biggr>\,,
\end{equation}
with the LHS colloquially denoting `the Feynman rule associated to the Lagrangian operator $\mathcal{O}_i = G_i f_i$.'  The simplest such objects are those where the Higgs dependence lies exclusively in the connections $G_i$, and where the momentum-dependence is trivial, e.g. the Majorana-mass and Yukawa-like interactions in \eqref{eq:MajoranaMassConnection}-\eqref{eq:YukawaConnection}.  For these couplings the all-orders Feynman rule is just an all-orders vertex function:\footnote{Recall that the mass-eigenstate Higgs coordinates are given by \eqref{eq:massbasisbosons}, and that SSB is realized by expanding $\phi_4$ around the vev, $\phi_4 \rightarrow \phi_4 + \overline{v}_T$.} 
 \begin{align}
 \label{eq:MajoranaRule}
 \lbrace \hat{h}, \overline{N}_p, N^c_r \rbrace = -i \,\biggl< \frac{\delta \eta_{N}(\phi)_{pr}}{\delta \hat{h}}\biggr> &= -i\,\sqrt{h}^{44}\, \sum_{n=0}^\infty \frac{(2n+2)}{2^{n+1}}\,\tilde{C}^{(5+2n)}_{\underset{pr}{NN}}\,\overline{v}_T^{2n+1} \,,\\
 \label{eq:YukawaRule1}
     \lbrace \hat{h}, \overline{N}_p, \ell_r \rbrace = -i \,\biggl< \frac{\delta \mathcal{Y}_{N}(\phi)_{pr}}{\delta \hat{h}}\biggr> &= i \frac{\sqrt{h}^{44}}{\sqrt{2}}Y^\dagger_{N,pr} - i \frac{\sqrt{h}^{44}}{\sqrt{2}}\sum_{n=0}^{\infty} \frac{(2n+3)}{2^{n+1}} \tilde{C}_{\underset{pr}{NH}}^{(6+2n)}\, \overline{v}_T^{2n+2}\,, \\
\label{eq:YukawaRule2}
     &= -i \sqrt{h}^{44} \frac{\overline{M}^D_{N,pr}}{\overline{v}_T} - i\frac{\sqrt{h}^{44}}{\sqrt{2}}\sum_{n=0}^{\infty} \frac{(2n+2)}{2^{n+1}} \tilde{C}_{\underset{pr}{NH}}^{(6+2n)}\, \overline{v}_T^{2n+2}\,,
 \end{align}
 where the first term in \eqref{eq:YukawaRule1} is simply the renormalizable Yukawa coupling from \eqref{eq:LagSS} and the vev $\overline{v}_T$ is defined as the minimum of the tree-level Higgs potentital in the (all-orders) geo($\nu$)SMEFT.  The Dirac mass matrix $\overline{M}^D_{N}$ in \eqref{eq:YukawaRule2} is simply the expectation value of the all-orders Yukawa connection, $\overline{M}^D_{N} \equiv \langle \mathcal{Y}_N(\phi) \rangle$.  Observe that the convention in \eqref{eq:YukawaRule1} (\eqref{eq:YukawaRule2}) is analogous to the geoSMEFT Yukawa rules convention found in \cite{Helset:2020yio} (\cite{Corbett:2021jox}).   

 A priori, with \eqref{eq:MajoranaRule}-\eqref{eq:YukawaRule2} (and its conjugate vertex) one can begin computing certain all-$\overline{v}_T/\Lambda$-orders amplitudes, e.g. for the tree-level $h \rightarrow \overline{N} \ell$ decay, or portions of the one-loop self-energy correction to the $N$ propagator.  However, extreme care must be taken when considering the Feynman rules of interacting Majorana fields due to additional subtleties in their associated Wick contractions with respect to purely Dirac particles.  A consistent formalism is presented in \cite{Denner:1992vza,Denner:1992me}, which introduces the notion of a \emph{fermion flow}, in addition to the standard \emph{fermion number flow} considered when only Dirac particles interact.  Furthermore, one must also take care when considering the implicit all-orders definitions of the Lagrangian parameters present in  \eqref{eq:YukawaRule1}, e.g. $\overline{M}^D_{N}$. For example, calculating amplitudes with definite flavors (e.g. e vs. $\mu$) requires knowledge of the fermion-mass-eigenstate basis and its associated flavor transformations, i.e. the PMNS matrix controlling leptonic charged currents.  We will discuss the former issue in future work, and briefly outline the challenges to treating neutrino flavor geometrically in the next Section.

\subsection{Neutrino Mass Eigenstates and Flavor Structures}
\label{sec:MASS&MIX}
The connections and operator forms enumerated in Sections \ref{sec:OPERATORS}-\ref{sec:CONNECTIONS} can be rotated to/from the mass-eigenstate basis of the weak gauge and/or Higgs bosons via \eqref{eq:bosonbasischange}-\eqref{eq:massbasisbosons},\footnote{Observe that these objects can be decomposed into \emph{tetrads} which simultaneously flatten the field-space metrics and diagonalize the asssociated mass matrices --- cf. the recent discussion in \cite{Cheung:2021yog}.  We see no reason why the rotations in fermion flavor space, composed of the flavor invariants we highlight in this Section, cannot also be understood via a tetrad decomposition, although this will also depend on any non-trivial metric that may be defined via (e.g.) fermion kinetic terms as discussed in the Summary Comments of Section \ref{sec:CONNECTIONS} above.} while rotating to the fermion mass-eigenstate basis, necessary for flavored phenomenology, introduces additional physical rotations into the Lagrangian. It is well known that (again see  \cite{Dasgupta:2021ies}) this basis change mixes active and sterile neutrino fields even at the renormalizable level, such that the gauge singlet can participate in weak interactions via its admixtures in active mass eigenstates.  The situation is even more complex in the geo($\nu$)SMEFT.

Specifically, organizing the operators above that contribute to the neutrino mass sector in the broken electroweak phase of the theory gives contributions of the form
\begin{equation}
\label{eq:massmatrix}
    \mathcal{L}_{\text{mass}} = 
    -\frac{1}{2}\left(\overline{\nu^c_L} \,\,\, \overline{N}\right) \cdot 
    \left(
    \begin{array}{cc}
   \langle \eta_\ell(\phi) \rangle & \langle  \mathcal{Y}^T_N(\phi)\rangle  \\
   \langle  \mathcal{Y}_N(\phi)\rangle & \langle \eta_N(\phi) \rangle 
    \end{array}
    \right) \cdot
    \left(
    \begin{array}{c}
    \nu_L \\
    N^c
    \end{array}
    \right) + \text{h.c.} \equiv  -\frac{1}{2} \overline{n} \, \mathcal{M}_\nu \, n + \text{h.c.}\,,
\end{equation}
 where we have used $ \overline{\nu^c_L} N^c = \overline{N} \nu_L$ and condensed flavor indices, such that $\left(\overline{\nu^c_L}, \overline{N} \right)$ is an $\left(n_\ell + n_f\right)$-dimensional row vector in flavor space, where $n_\ell \, (n_f)$ is again the number of SU(2)$_L$ doublet $\ell$ (SU(2)$_L$ singlet $N$) generations in the theory. Here the brackets $\langle \rangle$ indicate that the expectation value of the Higgs field has been taken in all of the field-space connections contributing to the tree-level mass terms.  The $\mathcal{Y}_N(\phi)$ and $\eta_N(\phi)$ objects are defined in \eqref{eq:YukawaConnection} and \eqref{eq:MajoranaMassConnection} respectively, and we have also introduced a novel `Weinberg Connection' $\eta_\ell(\phi)$ in the (1,1) entry of the flavor matrix in \eqref{eq:massmatrix}.  This object is the field-space connection built from the application of scalar dressings to the operator in \eqref{eq:SMEFTwithWeinberg}, i.e. the all-orders generalization of the $d=5$ Weinberg operator, and it has not yet been formally defined in the literature.\footnote{I thank Michael Trott for helpful discussions on this point.}  We do so via the following variation of the Lagrangian:
\begin{equation}
\label{eq:weinbergdefine}
    \eta_{\ell}(\phi)_{pr} \equiv \frac{\delta \mathcal{L}_{\text{SMEFT}}}{\delta (\overline{\ell^c_p} \ell_r)} \Big \vert_{\mathcal{L}(\alpha,\beta,..) \rightarrow 0} =\sum_{n=0}^\infty\left[ \tilde{H}^\dagger(\phi_I) \tilde{H}^\star(\phi_J) \, \tilde{C}_{\underset{pr}{\ell\ell}}^{(5+2n)}\right]\left(\frac{\phi^2}{2}\right)^n\,,
\end{equation}
 which obviously identities $\overline{\ell^c} \ell$ as the relevant composite two-point fermion bilinear operator in the unbroken (geo)SMEFT, and where in this convention each Higgs doublet is SU(2)$_L$-contracted with a lepton doublet in said bilinear.  There is also of course a Hermitian conjugate expression.

Returning to the issue of flavor, we recall that \eqref{eq:massmatrix} is readily diagonalized by a unitary transformation $U_n$ on the neutrino fields,
\begin{equation}
\label{eq:diagonalmass}
    \mathcal{U}_n^\dagger\, \mathcal{M}_\nu \, U_n \equiv m_\nu = \text{diag}\left(m_{\nu_1},...,m_{\nu_{n_l}},m_{N_1},...,m_{N_{n_f}} \right)\,,
\end{equation}
which amounts to a rotation to the neutrino mass-eigenstate basis of the Lagrangian.  The mass eigenvalues appearing on the RHS of \eqref{eq:diagonalmass}, and the mixing angles and CP-violating phases implicit in $U_n$, a priori exclusively depend on the connections appearing in \eqref{eq:massmatrix}.  In the spirit of writing down said Lagrangian parameters at all-orders in $\overline{v}_T/\Lambda$, one goal in the development of a complete tree-level geo$\nu$SMEFT should be their analytic derivation, which is achievable via invariant techniques along the lines of those presented in \cite{Talbert:2021iqn}, which builds on earlier work in flavor invariant and RG theory (in particular from \cite{Jenkins:2009dy,Feldmann:2015nia} and references therein) to extract exact, basis-independent formulae for calculating fermionic mass and mixing parameters exclusively as a function of field-space connections (at any order). 

For example, in the absence of a dynamical gauge-singlet $N$, one reverts to the geoSMEFT where tree-level neutrino masses are entirely described by the Weinberg Connection in \eqref{eq:weinbergdefine}, whose low-energy flavor structure is equivalent to the simplest type-I seesaw \cite{Minkowski:1977sc,Gell-Mann:1979vob,Mohapatra:1979ia,Yanagida:1980xy,Schechter:1980gr} models that leave only light, LH Majorana neutrinos in the IR spectrum.  The minimal basis of 15 flavor invariants necessary to unambiguously extract the physical parameters of this (three-generation) scenario was presented as early as \cite{Jenkins:2009dy}, while the complete basis of `primary' and `basic' invariants was presented in \cite{Wang:2021wdq} (also see \cite{Lu:2021yej,Lu:2021ada}), along with their RGE and a derivation of neutrino mass-eigenvalues and PMNS mixing angles and phases in terms of these invariants.  While \cite{Wang:2021wdq} did not consider a geometric EFT context, the flavor formulae presented there can be trivially extended to all orders by simply making the relevant invariants functions of $\eta_\ell(\phi)$, as opposed to a finite-order object.   

The limit where lepton-number violation is forbidden in \eqref{eq:massmatrix} is also simple, as the all-orders information encoded in $\mathcal{Y}_N(\phi)$ can be extracted with the formalism presented in \cite{Talbert:2021iqn} for the Dirac Yukawa sector of the (geo)SM(EFT).  The IR leptonic mass structure is analogous to that of the quarks --- the neutrino (charged lepton) mass eigenvalues are readily computed from three invariants composed exclusively of the $\mathcal{Y}_N(\phi)$ ($\mathcal{Y}_e(\phi)$) connections, while the (unitary contribution to the) PMNS mixing matrix parameterized by three real mixing angles and a lone Dirac CP-violating phase is computed from five additional invariants composed of \emph{both} $\mathcal{Y}_N(\phi)$  and $\mathcal{Y}_e(\phi)$. 

However, the situation is far more complex when the flavor structure of even the seesaw model ($\eta_\ell(\phi) = 0$) is considered,\footnote{...evaluated at an energy scale where $N$ are still propagating degrees of freedom.  Note that, while not impossible, it is perhaps less well-motivated theoretically to consider scenarios where $\eta_\ell(\phi)$ has $(n_f \times n_f)$ or more non-zero (flavor) matrix elements, as the standard picture is that non-zero $\tilde{C}_{\ell\ell}$ are induced upon integrating out a \emph{heavy} $N$, which would no longer contribute to structure in $\eta_{N}$ or $Y_N$.} much less the complete matrix in \eqref{eq:massmatrix}. As discussed in \cite{Jenkins:2009dy}, there are nine mass eigenvalues (three charged lepton terms $+$ six neutrino mass terms), six mixing angles, and six phases in the three-generation model, amounting to 21 physical parameters.  Naively this indicates that at least 27 flavor invariants are needed to unambiguously extract them. While the Hilbert Series for the three-generation seesaw model was eventually calculated using the Molien-Weyl formula in \cite{Hanany:2010vu}, finding the minimal basis of invariants implied by that (rather complex)  Hilbert Series remains to be done.\footnote{Note however that a complete basis was recently found for the minimal two-generation seesaw model in \cite{Yu:2021cco}.  Also see \cite{Yu:2022nxj,Yu:2022ttm} for more phenomenological considerations on CP violation and discussions regarding invariants in the `seesaw effective theory'.} Success would allow for the derivation of all mass, mixing, and CP-violating parameters implied in \eqref{eq:massmatrix} in terms of these invariants (and those associated to $\eta_\ell(\phi)$), relationships that would be flavor-basis-independent and hold at all-$\overline{v}_T/\Lambda$-orders, akin to the forms for the electroweak gauge bosons presented in \eqref{eq:allordergaugemass}, or the Dirac mass and (unitary) mixing parameters presented in \cite{Talbert:2021iqn}.

\section{Summary and Outlook}
\label{sec:CONCLUDE}
We have presented a geometric realization of the $\nu$SMEFT, the effective field theory defined by (non-)renormalizable operators invariant under Standard Model gauge symmetries and composed of only Standard Model fields and $n_f$ additional gauge-singlet (sterile) neutrinos $N$.  This geo$\nu$SMEFT allows for the definition of field-space connections valid at all-orders in $\overline{v}_T/\Lambda$, which contribute to the implied geometry of the Beyond-the-$\nu$Standard Model field space encoded by (non-)renormalizable operators in the standard $\nu$SMEFT. These connections are in many instances field-redefinition invariant and, at least at tree level, allow for the absorption of a tower of otherwise independent Wilson Coefficients into a single geometric object, thereby reducing the number of parameters in the relevant sector of the $\nu$SMEFT.  We have also briefly discussed the route to calculating all-orders amplitudes in the geo$\nu$SMEFT, including the invariant theory required to define the flavor parameters associated to its neutrino mass and mixing, as was achieved in \cite{Talbert:2021iqn} for the Dirac Yukawa sector of the geoSMEFT.  This required defining the 'Weinberg Connection' associated to the $d=5$ Weinberg operator and its all-orders generalization.  The formal development of the invariant theory for the full geo$\nu$SMEFT represents ongoing work \cite{MartinTalbert}.

In addition to  understanding the analytic flavor structure of the geo$\nu$SMEFT, and therefore the ability to calculate in its fermion mass-eigenstate basis, there are also larger open questions in the effort to understand bottom-up effective theories in geometric contexts.  For example, do all of the field-space connections $G_i$ maintain an all-$\overline{v}_T/\Lambda$-orders structure under renormalization group flow?  Is there another way of understanding the appearance of the $G_i$ from first principles, and what more can be learned about their role in defining and exploiting formal geometric quantities (e.g. fermionic metrics, curvature tensors, covariant derivatives...) on the geo($\nu$)SMEFT's field space and its associated scattering amplitude predictions?  Furthermore, what can be said about matching renormalizable theories to EFT amplitudes with all-orders $G_i$ dependence in tact?  The answers to these questions should be pursued in an effort to fully unlock the potential of the geometric approach to effective field theories.

\section*{Acknowledgements}
I thank Tyler Corbett, Andreas Helset, Adam Martin and Michael Trott for numerous pedagogical discussions regarding the (geo)SMEFT, Ilaria Brivio for her insight on neutrino physics, Bingrong Yu for his comments on invariant theory, as well as Coenraad Marinissen, Rudi Rahn, and Wouter Waalewijn for support with the {\tt{ECO}} package.  I also thank Adam and Andreas for their review of the manuscript, and Uli Haisch for his healthy skepticism of geometric EFTs.  I  gratefully acknowledge funding from the European Union's Horizon 2020 research and innovation programme under the Marie Sk\l{}odowska-Curie grant agreement No. 101022203, and also the hospitality and support of the Mainz Institute for Theoretical Physics, where portions of this project were completed.

\begin{appendix}
\begin{appendix}
\section{geoSMEFT Symmetry Generator Conventions}
\label{sec:CONVENTIONS}
In this section we recall some of the definitions of the symmetry generators presented above, as originally reported in \cite{Helset:2018fgq,Helset:2020yio}.  Critically, the $\Gamma_A$ matrices appearing in \eqref{eq:realconversion} are combinations of the $\gamma$ matrices also appearing therein,
\begin{equation}
\label{eq:GammaDefine}
    \Gamma_{A,K}^I = \gamma_{A,J}^I\,\gamma_{4,K}^J\,,
\end{equation}
where $\gamma$ are electroweak symmetry generators written in the real representation:
\begin{equation}
\gamma_{1,J}^I = \left(
    \begin{array}{cccc}
    0 & 0 & 0 & -1\\
    0 & 0 & -1 & 0\\
    0 & 1 & 0 & 0\\
    1 & 0 & 0 & 0\\
    \end{array}
    \right),\,\,\,
    \gamma_{2,J}^I = \left(
    \begin{array}{cccc}
    0 & 0 & 1 & 0\\
    0 & 0 & 0 & -1\\
    -1 & 0 & 0 & 0\\
    0 & 1 & 0 & 0\\
    \end{array}
    \right),\,\,\,
    \gamma_{3,J}^I = \left(
    \begin{array}{cccc}
    0 & -1 & 0 & 0\\
    1 & 0 & 0 & 0\\
    0 & 0 & 0 & -1\\
    0 & 0 & 1 & 0\\
    \end{array}
    \right),\,\,\,
    \gamma_{4,J}^I = \left(
    \begin{array}{cccc}
    0 & -1 & 0 & 0\\
    1 & 0 & 0 & 0\\
    0 & 0 & 0 & 1\\
    0 & 0 & -1 & 0\\
    \end{array}
    \right) \,.
\end{equation}
Inserting these expressions into \eqref{eq:GammaDefine} one quickly finds the following matrix representations for $\Gamma_A$:
\begin{equation}
    \Gamma_{1,J}^I = \left(
    \begin{array}{cccc}
    0 & 0 & 1 & 0\\
    0 & 0 & 0 & -1 \\
    1 & 0 & 0 & 0\\
    0 & -1 & 0 & 0\\
    \end{array}
    \right),\,\,\,
     \Gamma_{2,J}^I = \left(
    \begin{array}{cccc}
    0 & 0 & 0 & 1\\
    0 & 0 & 1 & 0 \\
    0 & 1 & 0 & 0\\
    1 & 0 & 0 & 0\\
    \end{array}
    \right),\,\,\,
     \Gamma_{3,J}^I = \left(
    \begin{array}{cccc}
    -1 & 0 & 0 & 0\\
    0 & -1 & 0 & 0 \\
    0 & 0 & 1 & 0\\
    0 & 0 & 0 & 1\\
    \end{array}
    \right),\,\,\,
    \Gamma_{4,J}^I = -\left(
    \begin{array}{cccc}
    1 & 0 & 0 & 0\\
    0 & 1 & 0 & 0 \\
    0 & 0 & 1 & 0\\
    0 & 0 & 0 & 1\\
    \end{array}
    \right)\,.
\end{equation}
In \eqref{eq:pointsform} we have also absorbed SU(2)$_L \times$U(1)$_Y$ gauge couplings into the Levi-Civita tensors and generator representations via
\begin{equation}
    \tilde{\gamma}_{A,J}^I = 
    \begin{cases}
    g_2\,\gamma_{A,J}^I,\,\,\,\text{for}\,\,\, A=1,2,3 \,, \\
    g_1\,\gamma_{A,J}^I,\,\,\,\text{for}\,\,\, A=4\,,
    \end{cases}
\end{equation}
\begin{equation}
    \tilde{\epsilon}^A_{BC} = g_2\,\epsilon^A_{BC}\,, \,\,\,\,\,\,\,\,\,\text{with}\,\,\,\,\,\,\,\,\, \tilde{\epsilon}^1_{23} = g_2\,,\,\,\,\,\,\tilde{\epsilon}^4_{BC}=0\,.
\end{equation}
Note that the generators $\gamma$ are defined in the weak-eigenstate basis, and can be transformed to their mass-basis representation via
\begin{equation}
\label{eq:gammaMassgen}
    {\bm{\gamma}}_{C,J}^I = \frac{1}{2} \tilde{\gamma}_{A,J}^I \sqrt{g}^{AB} U_{BC}\,.
\end{equation}
 Expanding indices in \eqref{eq:gammaMassgen} then gives the following  expressions for the mass-eigenstate SU(2)$_L\times$U(1)$_Y$ symmetry generators:
\begin{align}
\nonumber
    {\bm{\gamma}}_{1,J}^I &= \frac{\overline{g}_2}{2\sqrt{2}}\left(\gamma_{1,J}^I + i \gamma_{2,J}^I \right),\,\,\,\,\,\,\,\,
    {\bm{\gamma}}_{3,J}^I = \frac{\overline{g}_Z}{2}\left(c_{\theta_Z}^2 \gamma_{3,J}^I - s_{\theta_Z}^2 \gamma_{4,J}^I\right),\,\,\, \\
    {\bm{\gamma}}_{2,J}^I &= \frac{\overline{g}_2}{2\sqrt{2}}\left(\gamma_{1,J}^I - i \gamma_{2,J}^I\right),\,\,\,\,\,\,\,\,
    {\bm{\gamma}}_{4,J}^I = \frac{\overline{e}}{2}\left(\gamma_{3,J}^I + \gamma_{4,J}^I\right),
\end{align}
where the barred quantities are the all-orders generalizations of the mass-basis gauge couplings given in \eqref{eq:masscouplings} in the renormalizable ($\nu$)SM limit,
\begin{align}
    \overline{g}_2 &= g_2\sqrt{g}^{11} = g_2\sqrt{g}^{22} \,,\\
    \overline{g}_z &= \frac{g_2}{c^2_{\theta_Z}}\left(c_{\overline{\theta}} \sqrt{g}^{33}-s_{\overline{\theta}}\sqrt{g}^{34}\right) = \frac{g_1}{s^2_{\theta_Z}}\left(s_{\overline{\theta}} \sqrt{g}^{44}-c_{\overline{\theta}}\sqrt{g}^{34}\right)\,,\\
    \overline{e} &= g_2 \left(s_{\overline{\theta}} \sqrt{g}^{33}+c_{\overline{\theta}}\sqrt{g}^{34}\right) = g_1 \left(c_{\overline{\theta}} \sqrt{g}^{44}+s_{\overline{\theta}}\sqrt{g}^{34} \right)\,,
\end{align}
which are themselves functions of the all-orders generalizations of the weak mixing angles,
\begin{align}
    s^2_{\theta_Z} &=\frac{g_1\left(\sqrt{g}^{44}s_{\overline{\theta}}-\sqrt{g}^{34}c_{\overline{\theta}}\right)}{g_2 \left(\sqrt{g}^{33} c_{\overline{\theta}}-\sqrt{g}^{34}s_{\overline{\theta}}\right)+g_1\left(\sqrt{g}^{44}s_{\overline{\theta}}-\sqrt{g}^{34}c_{\overline{\theta}}\right)}\,,\\
    s^2_{\overline{\theta}} &= \frac{\left(g_1 \sqrt{g}^{44}-g_2\sqrt{g}^{34}\right)^2}{g_1^2\left[\left(\sqrt{g}^{34}\right)^2+\left(\sqrt{g}^{44}\right)^2\right]+g_2^2 \left[ \left(\sqrt{g}^{33}\right)^2+\left(\sqrt{g}^{34}\right)^2 \right]-2 g_1 g_2\sqrt{g}^{34}\left(\sqrt{g}^{33}+\sqrt{g}^{44}\right)}\,.
\end{align}
These are the quantities that appear in the (e.g.) geometric electroweak gauge boson masses found in \eqref{eq:allordergaugemass}.  
\end{appendix}
\end{appendix}

\bibliographystyle{unsrt}
\bibliography{biblionuSMEFT}

\end{document}